\documentclass[%
 reprint,
superscriptaddress,
nofootinbib,
 amsmath,amssymb,
 aps,
 showkeys,
]{revtex4-2}
\usepackage{verbatim}
\usepackage[english]{babel}
\usepackage{mathtools}
\usepackage{aps_macros}
\usepackage{natbib}
\usepackage{pifont}
\usepackage{graphicx}
\usepackage{dcolumn}
\usepackage{bm}
\usepackage{hyperref}
\usepackage{orcidlink}
\usepackage{subfigure}%
\usepackage{float}%
\hypersetup{
     colorlinks   = true,
     citecolor    = blue
}

\hypersetup{colorlinks=true, citecolor=blue, linkcolor = magenta}

\begin{document}

\preprint{APS/123-QED}

\title{Nonradial oscillations of realistic anisotropic neutron stars: Axial modes}

\author{Jose~F.~Rodriguez-Ruiz \orcidlink{0000-0003-3627-5084}} 
\email{jrodriguez154@uan.edu.co}
\affiliation{Departamento de F\'isica, Universidad Antonio Nari\~no, Cra 3 Este \# 47A - 15, Bogot\'a D.C. 110231, Colombia}

\author{L.~M.~Becerra \orcidlink{0000-0002-3262-5545}}
\email{laura.becerra@umayor.cl}
\affiliation{Centro Multidisciplinario de F\'isica, Vicerrector\'ia de Investigaci\'on, Universidad Mayor,  Santiago de Chile 8580745, Chile}
 

\author{F.~D.~Lora-Clavijo \orcidlink{0000-0003-4613-2917}} 
\email{fadulora@uis.edu.co}
\affiliation{Grupo de Investigaci\'on en Relatividad y Gravitaci\'on, Escuela de F\'isica, Universidad Industrial de Santander A. A. 678, Bucaramanga 680002, Colombia}

\date{\today}

\begin{abstract}
Non-radial oscillation modes of neutron stars serve as diagnostics of their internal composition and relativistic structure. In this work, we investigate the perturbations of static and spherically symmetric neutron stars characterized by an anisotropic pressure. Given the background symmetry, perturbations decouple into polar and axial modes. To date, axial modes have remained less explored, primarily because matter and metric perturbations decouple in the isotropic limit. In this work, we provide a consistent treatment of axial modes and demonstrate that pressure anisotropy induces a direct coupling between matter and metric perturbations. We employ parameterized anisotropy models that ensure consistency with the treatment of matter perturbations. We numerically integrate the linearized Einstein field equations for the axial modes, employing a diverse set of realistic equations of state. Our results indicate that as the stellar mass grows, the frequency of the base $w$-mode generally decreases, while its damping time increases. Softer equation of states typically yield slightly higher oscillation frequencies. Furthermore, larger anisotropy (i.~e.~, when the tangential pressure exceeds the radial pressure) allows for more massive equilibrium configurations, which correspondingly leads to lower oscillation frequencies and prolonged damping times. Finally, we demonstrate that the frequency and damping time, both scaled by the stellar mass,  exhibit a nearly universal quadratic dependence on the stellar compactness, remaining largely insensitive to  the underlying equation of state, but slightly dependent to the specific anisotropy model.
\end{abstract}

\maketitle


\section{\label{sec:intro} Introduction}

The investigation of neutron star  interiors requires a robust integration of General Relativity with the microphysics of matter at nuclear densities \cite{glendenning2012compact}. These compact objects serve as the premier astrophysical laboratories for probing the Equation of State (EOS) in regimes where the baryon number density $n_b$ far exceeds the nuclear saturation density \cite{2001ApJ...550..426L}. At these scales, the relation between  pressure and energy density  is governed by the strong interaction, and must be described through many-body calculations, typically based on Chiral Effective Field Theory or Relativistic Mean-Field models \cite{2013ApJ...773...11H}. To ensure a realistic description, the EOS must not only remain causality-preserving but also satisfy the stringent constraints imposed by multi-messenger observations. Specifically, the model must support the lower bound of $2.01 \pm 0.04 M_{\odot}$ for the maximum stellar mass observed in pulsars such as PSR J0348+0432 \cite{2013Sci...340..448A} and adhere to the tidal deformability limits derived from the gravitational wave event GW170817 \cite{2017PhRvL.119p1101A}. By grounding the stability and oscillation analysis in these realistic EOS, one can accurately map the internal composition of the star to its dynamic response, in particular to its non-radial oscillation modes, which provide a sensitive probe of high-density physics \cite{1998MNRAS.299.1059A}. 

Traditionally, neutron stars  have been modeled as isotropic fluids in hydrostatic equilibrium \citep{2000ARNPS..50..481H, 2016ARA&A..54..401O, 2016ApJ...820...28O}. However, various microphysical phenomena can generate significant pressure anisotropies within their high-density interiors, including ultra-strong magnetic fields \citep{2012MNRAS.427.3406F}, relativistic nuclear interactions \citep{1975ARA&A..13..335C}, and exotic phases such as interacting quark matter \citep{2019PhRvD.100j3006B}. Such deviations from isotropy fundamentally alter the stellar structure and its stability limits \citep{2026PhRvD.113b3009B}. Extensive studies, beginning with the foundational work of \citet{1974ApJ...188..657B}, demonstrate that anisotropy modifies the equilibrium configuration, influencing the mass-radius ratio \citep{2024PhRvD.109d3025B}, the moment of inertia \citep{2020EPJC...80..769R}, and oscillation frequencies. Recent research has further refined these models by investigating the pressure effects in realistic anisotropic stars \citep{2024PhRvD.109d3025B} and expanding the framework to include slowly rotating configurations with parametrized equations of state \citep{2024PhRvD.110b4052B,2024PhRvD.110j3004B}. Moreover, comparative analyses of different anisotropy models in rotating stars have highlighted how internal stress distributions directly impact observable macroscopic properties \citep{2025PhRvD.111j3005B}.

The inclusion of anisotropic pressure is particularly critical when reconciling theoretical models with multi-messenger observational data. Recent studies have used gravitational wave data from the GW170817 and GW190814 events to constrain the range of pressure anisotropy and its impact on tidal deformability and $f$-mode frequencies \citep{2019PhRvD..99j4002B, 2022PhRvD.106j3518D, 2023arXiv230515724R}. Furthermore, mass and radius measurements from the NICER observations of pulsars PSR J0030+045 \citep{2019ApJ...887L..21R, 2019ApJ...887L..24M} and PSR J0740+662 \citep{2021ApJ...918L..27R, 2021ApJ...918L..28M} provide tighter bounds on the permissible anisotropy within a realistic neutron star. Modeling these effects is essential for understanding the behavior of matter under extreme conditions, as anisotropy can stabilize stellar configurations that would otherwise be deemed unstable under the assumption of local isotropy \citep{2011CQGra..28b5009H, 2018EPJC...78..673E}.   

The dynamical response of neutron stars to perturbations provides a unique window into their internal composition, a field known as gravitational-wave asteroseismology \citep{1999LRR.....2....2K}. Non-radial oscillations are characterized by a spectrum of discrete quasi-normal modes, which are traditionally classified according to their restorative forces: the fundamental $f$-mode, pressure $p$-modes, and buoyancy-driven gravity $g$-modes \citep{1941MNRAS.101..367C}. In addition to these fluid modes, the general relativistic framework predicts the existence of spacetime $w$-modes, which represent pure gravitational-wave degrees of freedom \citep{1992MNRAS.255..119K}. While the $f$-mode frequency is closely correlated with the average density of the star, the $w$-modes are sensitive to the curvature of spacetime and the stellar compactness. In the presence of realistic anisotropic pressure distributions, the characteristic frequencies and damping times of the polar modes are expected to deviate from those calculated in the Cowling approximation, which neglects spacetime fluctuations \citep{2012PhRvD..85l4023D, 2022PhRvD.106j3518D}.
Although recent works by \citet{2024PhRvD.110h3020L} established a consistent general relativistic framework for the non-radial polar (even-parity) perturbations of anisotropic neutron stars, the axial (odd-parity) sector has been poorly  unaddressed. Accurate modeling of these perturbations is essential for developing signal templates for third-generation detectors, such as the Einstein Telescope and Cosmic Explorer, which aim to resolve the high-frequency spectra of post-merger remnants \citep{2021PhRvC.104f5805R, 2023arXiv230515724R}. 

To the best of our knowledge, axial modes have mainly been studied in the context of black holes (see, e.g., \cite{Zhang:2021bdr, Zhao:2023tyo, Konoplya:2024lch}), where an effective anisotropic energy momentum tensor is introduced. However, in those treatments, the perturbation of the normal space vector is taken as zero, decoupling the  matter and metric sectors. Therefore, the anisotropy only indirectly affects the metric perturbation via a modified Zerilli potential.  Besides, axial modes have been also studied for isotropic relativistic stars in \cite{1991RSPSA.434..449C, 1994MNRAS.268.1015K, Benhar:1998au, Blazquez-Salcedo:2012hdg}. In that case, the axial metric perturbations do not excite any fluid motion, in contrast with the scenario studied in this work. On the other hand, the equations governing axial perturbations for a non-perfect fluid have been studied in \citep{Diaz-Guerra:2024gff}, but given their complexity, they have not been solved. More recently, in \cite{Bussieres:2026rnz}, axial modes have been studied as driven by viscosity and using a polytropic EOS.

In this work, we extend the perturbation analysis to include the axial case, where the perturbations are characterized by purely rotational fluid displacements and spacetime curvature fluctuations. Unlike the polar case, axial perturbations in non-rotating stars do not involve variations in the energy density or pressure; however, they admit a distinct spectrum of spacetime $w$-modes. By solving the linearized axial Einstein field equations, we provide the first complete description of the axial quasi-normal mode spectrum for realistic anisotropic neutron stars, thereby filling a significant gap in the current relativistic asteroseismology literature.

This work is organized as follows. In Sec. \ref{sec:Pert}, we obtain the perturbation equations of the axial modes and detail the realistic equations of state alongside the specific pressure anisotropy prescriptions (the Bowers-Liang and Horvart models) utilized in our framework. In Sec. \ref{sec:numerical}, we describe the numerical implementation, outlining the integration scheme, the handling of the singular stellar surface, and the matching of boundary conditions required to compute the complex quasinormal mode frequencies. Section \ref{sec:axialmodes} presents our main results for the axial $w$-modes, analyzing how their oscillation frequencies and damping times vary with stellar mass and anisotropy, and introducing empirical universal relations dependent on stellar compactness. Finally, in Sec. \ref{sec:discussion}, we provide our discussions and concluding remarks. Explicit expressions for the static background structure and the axial perturbation equations, including their necessary series expansions near the origin, are compiled in Appendix \ref{app:I}.

\section{\label{sec:Pert} Perturbations of static and spherically symmetric spacetime}

The background spacetime for a static, spherically symmetric neutron star is described by the line element in Schwarzschild coordinates $(t, r, \theta, \phi)$:
\begin{equation}
    ds^2 = -A(r)dt^2 + B(r)^{-1}dr^2 + r^2(d\theta^2 + \sin^2\theta d\phi^2).
\end{equation}
To investigate non-radial oscillations, we introduce a linear perturbation $h_{\alpha\beta}$ to the background metric $\bar{g}_{\alpha\beta}$:
\begin{equation}
    g_{\alpha\beta} = \bar{g}_{\alpha\beta} + h_{\alpha\beta}, \quad |h_{\alpha\beta}| \ll |\bar{g}_{\alpha\beta}|.
\end{equation}
All barred variables denote background fields, which are an equilibrium solution of the Einstein field equations.
Exploiting the spherical symmetry of the background, the perturbations are expanded in terms of spherical tensor  harmonics. 
The perturbations decouple into two independent subsets based on their parity: axial (odd-parity) and polar (even-parity):
\begin{equation}
    h_{\alpha\beta} = \sum_{\ell=0}^{\infty}\sum_{m=-\ell}^{m=\ell} (h^{\ell m}_{\alpha\beta})^{\rm axial} + (h^{\ell m}_{\alpha\beta})^{\rm polar}.
\end{equation}
The unperturbed NS is described by an anisotropic fluid  whose energy-momentum tensor (EMT) is
\begin{equation}
    \bar{T}_{\alpha \beta} = (\bar{\epsilon} + \bar{P}_\perp) \bar{u}_\alpha \bar{u}_\beta
    + \bar{P}_\perp \bar{g}_{\alpha \beta} + (\bar{P} - \bar{P}_\perp) \bar{k}_\alpha \bar{k}_\beta, \label{eq:T_backg}
\end{equation}
where $\bar{\epsilon}$ is the energy density, and $\bar{P}$ and $\bar{P}_\perp$ are the radial and tangential pressures, respectively. The fluid four-velocity is the normalized timelike vector $u^\alpha = \delta^\alpha_t / \sqrt{A}$, satisfying $u_\alpha u^\alpha = -1$. Correspondingly, $k^\alpha = \delta^\alpha_r / \sqrt{B}$ is a unit spacelike radial vector that is transverse to the flow, such that the orthogonality condition $u_\alpha k^\alpha = 0$ is satisfied.

We consider a perturbation of the neutron star where each fluid element undergoes a displacement defined by the Lagrangian vector field $\xi^\alpha$.This displacement modifies the fluid properties. The Lagrangian perturbation,  $\Delta$, is related to the Eulerian perturbation,  $\delta$, through the Lie derivative along the displacement vector: 
\begin{equation}
\Delta = \delta + \mathcal{L}_\xi, \label{eqn:Lagragian-Eulerian-pert}
\end{equation}
The geometric constraints, specifically the normalization of the four-velocity $u^\alpha$ and the space-like vector $k^\alpha$, along with their orthogonality, must remain preserved under this perturbation:
\begin{equation}
\Delta (g_{\alpha\beta }u^\alpha u^\beta) = 0, \label{eqn:Deltau2}
\end{equation}
\begin{equation}
\Delta(g_{\alpha\beta }k_\alpha k^\beta) = 0, \label{eqn:Deltak2}
\end{equation}
\begin{equation}
\Delta (g_{\alpha\beta }u^\alpha k^\beta) = 0. \label{eqn:Delatuk}
\end{equation}
Imposing the normalization condition defined in Eq. \eqref{eqn:Deltau2} allows us to express the Lagrangian perturbation of the four-velocity in terms of the metric variation as follows:
\begin{equation}
\Delta u^\alpha = \frac{1}{2}u^\alpha u^\mu u^\nu \Delta g_{\mu\nu}. \label{eqn:Deltau}
\end{equation}
In this work, we take the fluid displacement vector $\xi^\mu$  as the fundamental dynamical variable for perturbations of the matter sector. We assume that the anisotropy is an intrinsic property of each fluid element, it is frozen into the element and Lie-dragged along with it under displacement. Physically, this means that a displaced fluid element retains its anisotropy axis, which is simply carried along by the flow without acquiring any independent dynamics. Since $k^\alpha$ is defined geometrically as the unit radial vector of the background spherical symmetry, its Lagrangian perturbation is fully determined by the metric perturbation and the displacement $\xi^\mu
$. As a consequence, $\Delta k^\alpha$
lies in the $(u,k)$ plane. Therefore, from Eqs.~\eqref{eqn:Deltak2} and \eqref{eqn:Delatuk} the Lagrangian perturbation of $k^\alpha$ is given by,
\begin{equation}
    \Delta k^\alpha = u^\mu k^\nu \Delta g_{\mu\nu} u^\alpha - \frac{1}{2} k^\mu k^\nu \Delta g_{\mu\nu} k^\alpha. \label{eqn:Deltak}
\end{equation}
The first term restores orthogonality between $k^\alpha$ and $u^\alpha$ in the perturbed geometry, while the second term restores the unit normalization of $k^\alpha$. Together they ensure that the geometric constraints are satisfied to first order under any metric perturbation.

The Eulerian perturbation can be easily obtained from Eq.~\eqref{eqn:Lagragian-Eulerian-pert}. It is worthwhile to mention that the perturbations of the covariant components of four-velocity and the spacelike radial vector are not related to the perturbation of their contravariant components by simply the background metric. They are related as follows
\begin{align}
    \delta u_\alpha &= \bar{g}_{\alpha\beta}\delta u^\beta + h_{\alpha\beta}\bar{u}^\beta,\\
    \delta k_\alpha &= g_{\alpha\beta}\delta k^\beta + h_{\alpha\beta}\bar{k}^\beta.
\end{align}
The Eulerian perturbation of the EMT is
\begin{multline}
\delta T_{\mu\nu} = (\bar{\epsilon} + \bar{P}_{\perp})(\delta u_{\mu}\bar{u}_{\nu} + \bar{u}_{\mu}\delta u_{\nu}) \\
+ (\bar{P} - \bar{P}_{\perp})(\delta k_{\mu}\bar{k}_{\nu} + \bar{k}_{\mu}\delta k_{\nu}) \\
+ \bar{P}_{\perp}h_{\mu\nu} + (\bar{g}_{\mu\nu} + \bar{u}_{\mu}\bar{u}_{\nu} - \bar{k}_{\mu}\bar{k}_{\nu})\delta P_{\perp} \\
+ \bar{u}_{\mu}\bar{u}_{\nu}\delta\epsilon + \bar{k}_{\mu}\bar{k}_{\nu}\delta P,
\end{multline}
and the perturbed Einstein equations are:
\begin{equation}
    \delta G_{\mu\nu} = 8\pi \delta T_{\mu\nu}.
\end{equation}
As usual, we analyze the perturbation in the Regge-Wheeler gauge. \citep{Regge:1957td}. Since the background is invariant under rotations, we can assume that the perturbations are axisymmetric, i.e., we set $m=0$ in all the multipolar expansions\footnote{Due to the spherical symmetry of the background, perturbations with azimuthal dependence can be derived from their axisymmetric counterparts through an appropriate rotation of the coordinate system; see \cite{chandrasekhar1998mathematical} for a detailed discussion.}.

Finally, in this work we employ the Bowers-Liang and Horvart anisotropy models (see section~\ref{sec:anistropy-model}). We emphasize that these  models are phenomenological prescriptions defined on the equilibrium background and carry no specification of how the anisotropy responds dynamically to fluid displacement. In the absence of a microscopic model, we adopt the frozen-in assumption as the minimal closure: each fluid element Lie-drags its anisotropy along its worldline. This is the simplest assumption consistent with covariance and closes the perturbation system without additional parameters. We leave the study of alternative closures for future work.

\subsection{Axial Perturbations}

The axial perturbations of the metric are 
\begin{equation}
(h^{\ell 0}_{\alpha\beta})^{\rm axial}=
\begin{bmatrix}
0 & 0 & 0 & h_0 \\
0 & 0 & 0 & h_1 \\
0 & 0 & 0 & 0 \\
h_0 & h_1 & 0 & 0
\end{bmatrix}
\left( \sin\theta \partial_\theta\right) Y^{\ell0}(\theta),  \label{eqn:h_axial}
\end{equation}
where $h_0$ and $h_1$ are functions of $t$ and $r$. Regarding the matter sector, the corresponding axial Lagrangian displacement vector is defined as:
\begin{equation}
    (\xi^\alpha)^{\rm axial}=[0,0,0, X (\sin\theta)^{-1} \partial_\theta Y_{\ell0}],
\end{equation}
where $X$ is a function of $t$ and $r$. Now, from Eqs \eqref{eqn:Lagragian-Eulerian-pert} and \eqref{eqn:Deltak} we obtain the Eulerian perturbations,
\begin{equation}
    (\delta u^\alpha)^{\rm axial} =[0,0,0, (\sqrt{A} \sin\theta)^{-1} \partial_\theta Y_{\ell0}\partial_t X],
\end{equation}
\begin{equation}
    (\delta k^\alpha)^{\rm axial} =[0,0,0,
\sqrt{B}(\sin\theta)^{-1} \partial_\theta Y_{\ell0}\partial_rX].
\end{equation}

The background energy density, radial pressure, and tangential pressure are scalars under the rotation group on the two-sphere; consequently, they do not provide source terms for axial perturbations in the isotropic case. In the presence of anisotropy, however, the perturbed EMT reduces to:
\begin{equation}
(\delta T^{\ell 0}_{\alpha\beta})^{\rm axial}=
\begin{bmatrix}
0 & 0 & 0 & s^{Bt} \\
0 & 0 & 0 & s^{B1} \\
0 & 0 & 0 & 0 \\
s^{Bt} & s^{B1} & 0 & 0
\end{bmatrix}
\left( \sin\theta\, \partial_\theta\right) Y^{\ell0}(\theta), \label{eqn:dT_axial} 
\end{equation}
where the axial source functions $s^{Bt}$ and $s^{B1}$ are defined as:
\begin{equation}
    s^{Bt} = - r^2(\bar{\epsilon} + \bar{P} -\bar{\sigma})\partial_tX-\bar{\epsilon}\,h_0,
\end{equation}
\begin{equation}
    s^{B1} = r^2\bar{\sigma}\,\partial_rX + \bar{P}\, h_1,
\end{equation}
with the anisotropy parameter $\bar{\sigma}=\bar{P}-\bar{P}_\perp$. After the substitution of Eqs. \eqref{eqn:h_axial} and \eqref{eqn:dT_axial} into the perturbed Einstein field equations, we obtain
\begin{equation}
\begin{split}
    &\frac{1}{2} \left[ B \left( \frac{4}{r} - \frac{A'}{A} \right) + B' \right] \partial_t h_1 + B \partial_t \partial_r h_1 \\
    &+ \frac{1}{2} \left( \frac{B A'}{A} - B' \right) \partial_r h_0 - B \partial_r^2 h_0 \\
    &+ \frac{1}{2} \left( \frac{2 B A'' + A' B'}{A} - \frac{B A'^2}{A^2} \right) h_0 \\
    &+ \frac{ \left( 2 r B' + 2 B + \ell^2 + \ell - 2 \right) }{r^2} h_0 = 16\pi s^{Bt},
\end{split}
\end{equation}
\begin{equation}
\begin{split}
    &\frac{2}{r} \partial_t h_0 - \partial_t \partial_r h_0 + \partial_t^2 h_1 + \frac{B ( r A'' + A' )}{r} h_1 \\
    &+ \frac{1}{2} A' B' h_1 - \frac{B A'^2}{2 A} h_1 \\
    &+ \frac{A ( r B' + \ell^2 + \ell - 2 )}{r^2} h_1 = 16\pi A s^{B1},
\end{split}
\end{equation}
\begin{equation}
    \frac{1}{2} (B A' + A B') h_1 + A B \partial_r h_1 - \partial_t h_0 = 0.
\end{equation}

The perturbation of the conservation of the EMT  $\delta(\nabla_\alpha T^{\alpha\beta})=0$, gives an equation for $X$,
\begin{multline}
       \left(\frac{A' }{2A}+\frac{B' }{2B} +\frac{2}{r}\right)\bar{\sigma}B(h_1 + r^2\partial_rX) \\
       +\partial_r[\bar{\sigma}B(h_1 + r^2\partial_rX) ]\\
       +\frac{ (\bar{P}+\bar{\epsilon}-\bar{\sigma})}{A}\left( \partial_t h_0 + r^2\partial^2_tX\right)=0.
\end{multline}
For a perfect-fluid background, this equation reduces to: $\partial_t(h_0 + r^2\partial_tX)=0$, implying that $\partial_t\delta u_\mu=0$.
This equation can be derived from the perturbed Einstein equations, thus there are only three independent equations for three unknown functions.

The master equation for the axial perturbations, analogous to the Regge-Wheeler  equation for the axial modes, is expressed in terms of the tortoise coordinate $x$ as:
\begin{equation}
(\partial_t^2-\partial_x^2 + V_{\rm axial})\Psi^{\rm axial}= S_{\rm axial},
\end{equation}
where the tortoise coordinate is defined by $dx/dr = 1/\sqrt{AB}$, and the master function is given by $\Psi^{\rm axial} = \sqrt{AB} \, h_1 / r$. The effective potential $V_{\rm axial}$ accounts for the anisotropic nature of the background fluid:
\begin{equation}
V_{\rm axial} = \frac{A\left[\ell (\ell+1) - 2 + 2 B\right]}{r^2} - \frac{(AB)'}{2 r} - 16\pi A\, \bar{\sigma}.
\end{equation}
In the vacuum, this potential reduces to the standard Regge-Wheeler potential for a Schwarzschild background ($A = B = 1 - 2M/r$):
\begin{equation}
V_{\rm axial}^{\rm Schw} = \left(1-\frac{2M}{r}\right)\left[\frac{\ell(\ell+1)}{r^2}-\frac{6M}{r^3}\right].
\end{equation}
The source term $S_{\rm axial}$, which couples the metric perturbations to the fluid displacement, is defined as:
\begin{equation}
S_{\rm axial} = 16 \pi A r \bar{\sigma}  \partial_r X.
\end{equation}
This source term vanishes for an isotropic background fluid, reflecting the fact that axial modes do not couple to the matter sector in the absence of anisotropy.

\subsection{Equation of state}

We select three EOS that represent distinct particle populations and support maximum masses exceeding $2M_\odot$, consistent with current observational constraints \cite{2019ApJ...887L..21R,2022ApJ...934L..17R}. These include: the SLy4 EOS, describing purely nucleonic matter; the GM1Y6 EOS, which incorporates hyperonic degrees of freedom; and the QHC21 EOS, a hybrid model accounting for nucleons, hyperons, and desconfined quark matter \footnote{The QHC21 EOS belongs to the  Quark–Hadron Crossover family, where the transition from hadronic to quark matter occurs smoothly rather than through a sharp first-order phase transition.}.

These three EOSs are parametrized using the Generalized Piecewise Polytropic (GPP) scheme introduced by \citet{Boyle2020}. In this approach, the rest-mass density is divided into $N$ intervals. Within each interval, $\rho_i \leq \rho \leq \rho_{i+1}$, the pressure $P$ and energy density $\epsilon$ are defined by
\begin{eqnarray}\label{eq:poly_eos}
    P(\rho) &=& K_i\rho^{\Gamma_i} + \Lambda_i \,,\\
    \epsilon(\rho) &=& \frac{K_i}{\Gamma_i-1}\rho^{\Gamma_i} +(1+a_i)\rho - \Lambda_i \,, 
\end{eqnarray}
where polytropic indices, $\Gamma_i$ together with  the dividing densities $\rho_i$ define the  parametrization. The remaining coefficients $K_i$, $\Lambda_i$, and $a_i$ are determined by ensuring the continuity of the energy density, pressure, and sound speed at the dividing densities. 

For the high-density regime, defined by densities exceeding the nuclear saturation density, $\rho_s \approx 2.4 \times 10^{14}$ g cm$^{-3}$, we employ a three-zone GPP model. Conversely, for the low-density region ($\rho < \rho_s$), we adopt a five-zone parameterization \citep[see][for details and specific parameter values]{PhysRevD.109.043025}.
 

\subsection{Anisotropy models}\label{sec:anistropy-model}
In this study, we adopt two widely utilized prescriptions for pressure anisotropy, the Bowers-Liang model \citep{1974ApJ...188..657B} and the Horvat model \citep{PhysRevD.85.124023,PhysRevD.109.043025}:
\begin{eqnarray}
\bar{\sigma}_{BL} &=& -\lambda_{BL} (\bar{\epsilon} + 3\bar{P})(\bar{\epsilon} + \bar{P}) \frac{r^2}{B} , \label{eq:sigma_BL}\\
\bar{\sigma}_H    &=&  -\lambda_H  \left(1-B\right) \bar{P} \, ,\label{eq:Delta_p}
\end{eqnarray}
where $\lambda_{BL}$ and $\lambda_H$ are dimensionless parameters that modulate the magnitude of anisotropy within the 
star. Following \citep{2025PhRvD.111j3005B}, we restrict these constants to the ranges $0 \leq \lambda_{BL} < 1$ and $0 \leq \lambda_{H} < 1$. These constraints ensure a monotonically decreasing radial pressure profile and guarantee that the tangential sound speed remains finite at the stellar center.

We should clarify that the anisotropy models in Eqs.~(\ref{eq:sigma_BL}) and~(\ref{eq:Delta_p}) are phenomenological. They are written directly in terms of the radial and tangential pressure components in the fluid rest frame, rather than as manifestly covariant tensorial relations. This is a common practice in the literature on anisotropic neutron stars, where such algebraic prescriptions are typically imposed in Schwarzschild coordinates to close the system of structure equations \citep{PhysRevD.85.124023,2025PhRvD.111j3005B,2024PhRvD.110h3020L,Mondal:2026hki}. Strictly speaking, these models are not covariantly defined in arbitrary coordinate systems. However, two aspects mitigate this concern in our specific context. First, since we work with static, spherically symmetric backgrounds, the areal radius $r$ is a geometric invariant, not a mere coordinate. Second, and more importantly, axial perturbations do not change the radial structure of the star at linear order. The background quantities $\bar{\epsilon}$, $\bar{P}$ and $\bar{P}_\perp$ remain unperturbed, since they are scalars under rotations. The only fluid quantities that change are the four-velocity $u^\alpha$ and the transverse vector $k^\alpha$, whose variations are dictated by the covariant perturbation relations. Thus, the fact that the background anisotropy models are not manifestly covariant has a limited impact on the axial oscillation modes we study.

\section{Numerical implementation}\label{sec:numerical}
\begin{figure*}
    \centering
    \includegraphics[width=0.99\linewidth]{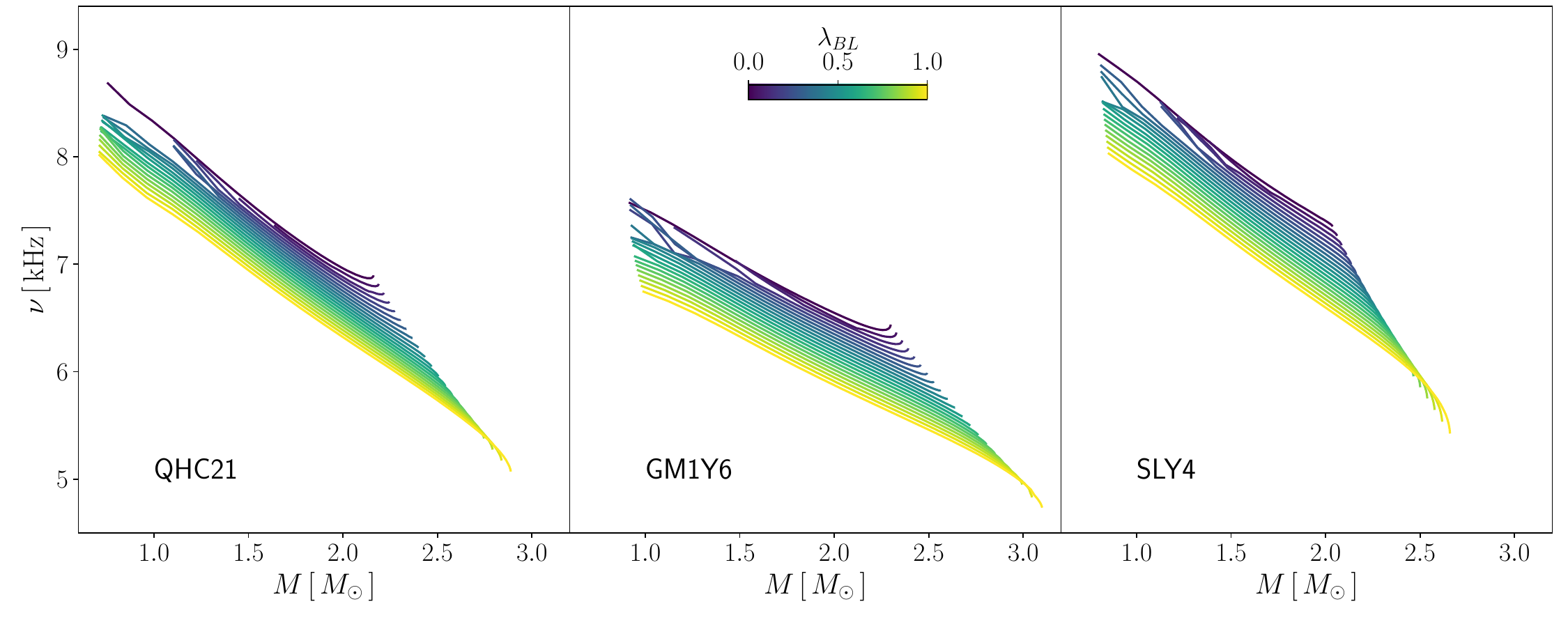}
    \includegraphics[width=0.99\linewidth]{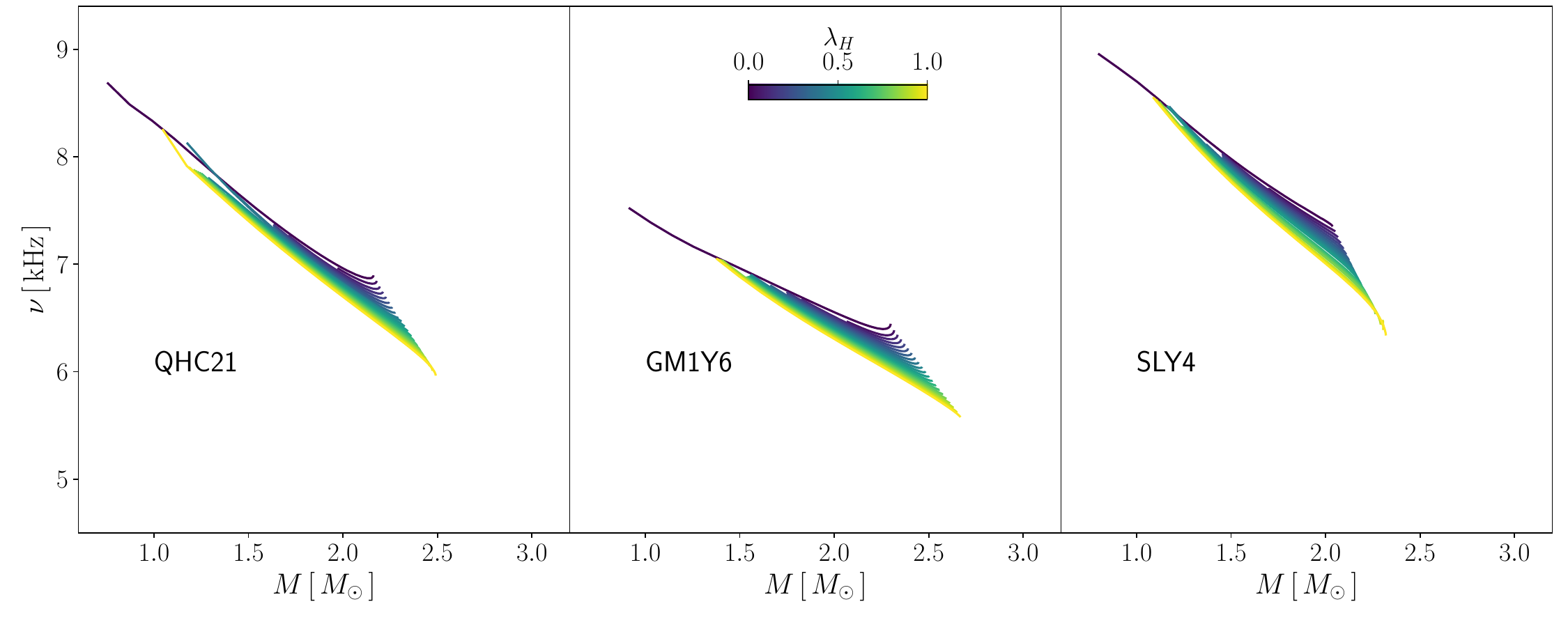}
    \caption{Oscillation frequency of the base $w$-mode as a function of the stellar mass for different EOS and the Bowers-Liang (upper panel) and the Horvart (lower panel) anisotropy model. The color scale corresponds to the value of the anisotropic parameter. }
    \label{fig:fvsM}
\end{figure*}

\begin{figure*}
    \centering
    \includegraphics[width=0.99\linewidth]{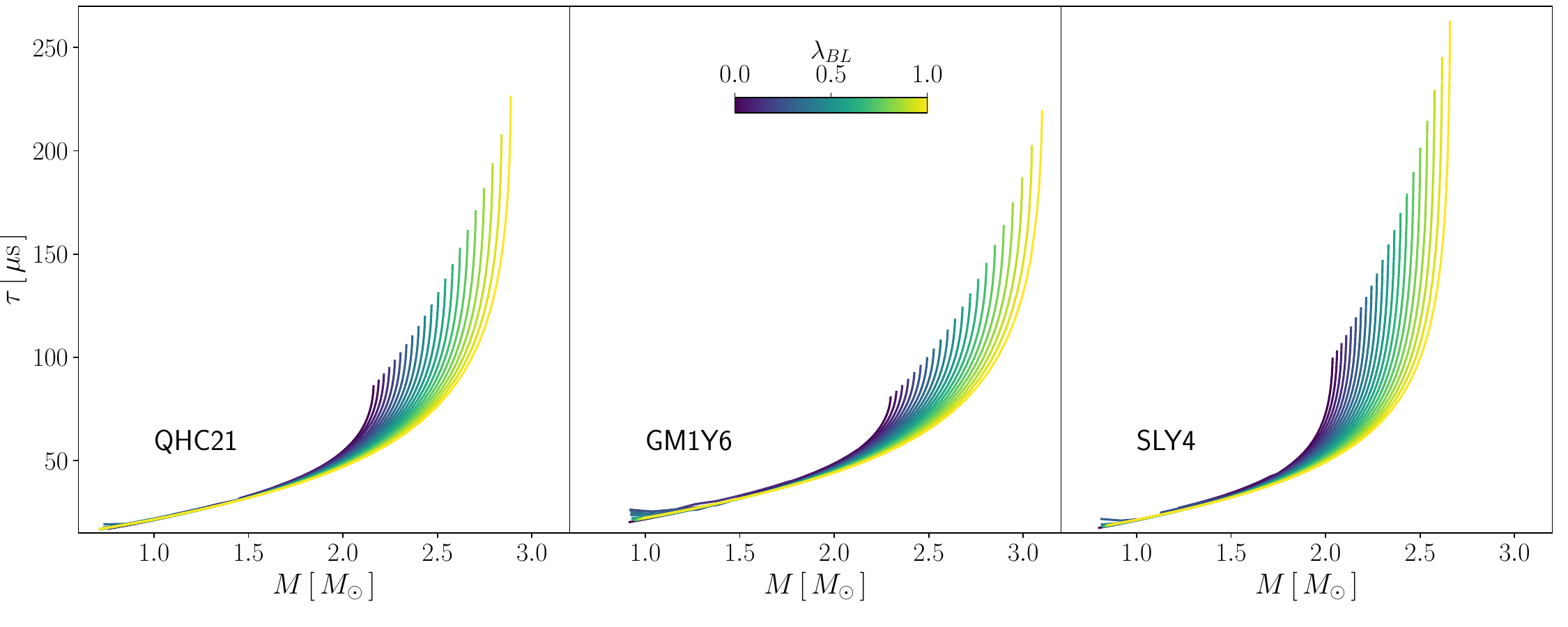}
    \includegraphics[width=0.99\linewidth]{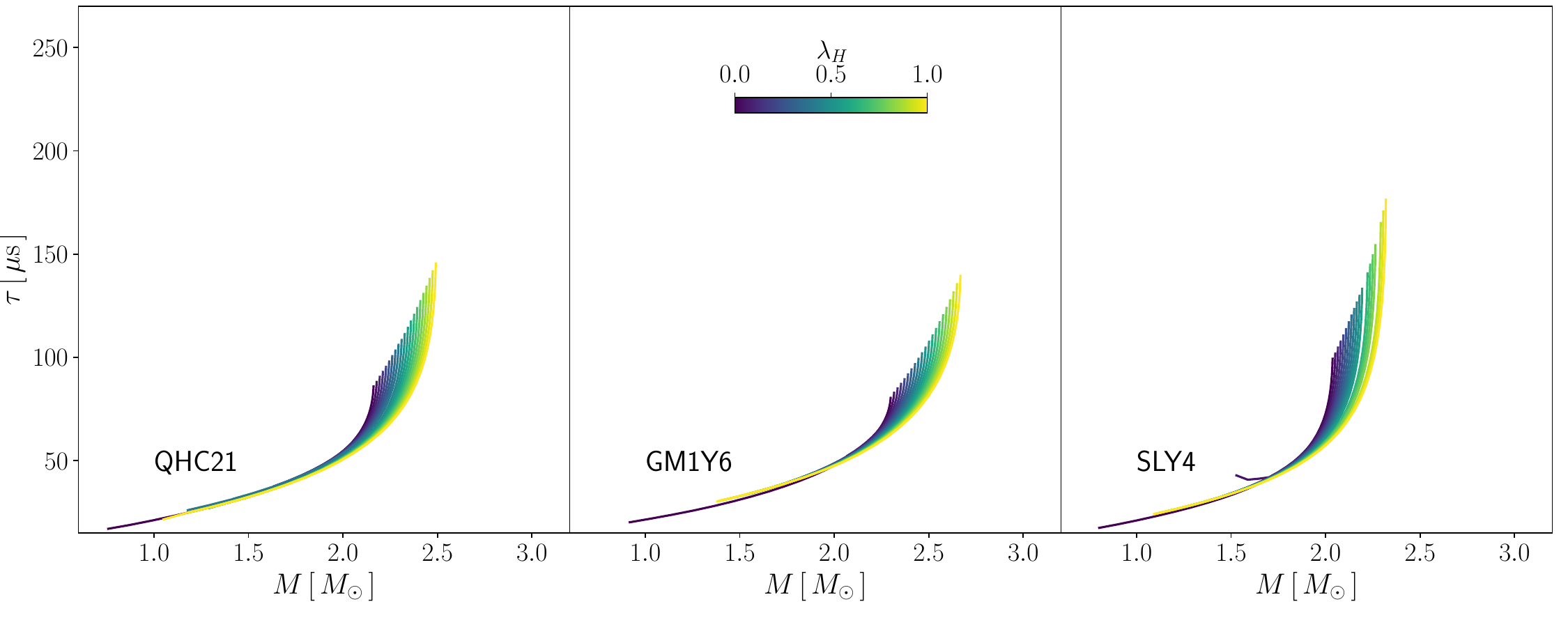}
    \caption{Same as Figure~\ref{fig:fvsM}  but for the damping time as a function of the stellar mass. }
    \label{fig:tauvsM}
\end{figure*}

The complete system of equations for numerical integration is provided in Appendix~\ref{app:I}. The static, anisotropic background is governed by equations \eqref{eq:da}–\eqref{eq:dP}, which recover the Schwarzschild vacuum solution ($A = B = 1 - 2M/r$) outside the stellar surface, where $P = P_\perp = \epsilon = 0$. For the axial sector, we adopt a harmonic time dependence of the form $e^{i\omega t}$ for all perturbed quantities, with the dynamics described by Eqs. \eqref{eq:axiala}–\eqref{eq:axiald}. The quasinormal modes are identified as the discrete set of complex frequencies, $\omega = \omega_R + i\omega_I$, that satisfy specific physical boundary conditions: the solutions must remain regular at the coordinate origin, while the perturbed metric functions, $h_0$ and $h_1$, must behave as purely outgoing waves in the far-field limit ($r \to \infty$).

All differential equations are integrated using a fourth-order Runge-Kutta integrator in a 1D spherical grid extending from $r = 0M$ to the outer domain boundary, $r_{\rm max} = 50 M$.   To compute the mode frequencies of the pulsating star, we proceed as follows:
 \begin{itemize}
 \item First, setting the central value of the energy density ($\bar{\epsilon}(r=0)=\bar{\epsilon}_c$), we integrate the background equations \eqref{eq:da}–\eqref{eq:dP} together with the perturbation equations \eqref{eq:axiala}–\eqref{eq:axiald} from the origin to the stellar surface. To ensure regularity at the origin, we construct series expansions of the solutions near $r=0$ (see Appendix~\ref{app:I}) and use them to initialize the integration at the first grid point, $r=\Delta r$, where $\Delta r$ is the uniform radial step. The integration then proceeds outward up to the stellar surface $R_s$, defined by $\bar{\rho}(R_s)= 10 ^4$~ g cm$^{-3}$.
 
 The adopted anisotropic prescriptions satisfy $\sigma(r=R_s)=0$, which causes the perturbation equations to become singular at the surface. To handle this, we construct two independent solutions of the perturbation equations, $\psi^{(1)}\equiv \{h_0^{(1)}, h_1^{(1)}, \chi^{(1)}, X^{(1)}\}$ and $\psi^{(2)}\equiv \{h_0^{(2)}, h_1^{(2)}, \chi^{(2)}, X^{(2)}\}$, and write the interior solution as their linear combination, $\psi_{(\mathrm{in})} = \psi^{(1)} + D,\psi^{(2)}$. The constant $D$ is then fixed by imposing the condition $\chi(R)=0$.

\item Next, we compute the exterior solution $\psi_{\rm (out)}$ of the perturbed equations, by integrating the Regge-Wheeler equation in the vacuum exterior of the star, where $P = P_\perp = \epsilon = 0$ and the background spacetime is described by the Schwarzschild metric with mass $M=(1-B(R))R/2$. The integration is carried out from the outer boundary of the grid $r_{\rm max}$, inward toward the stellar surface $R$. To impose the appropriate boundary condition, we initialize the integration at $r_{\rm max}$ using the asymptotic outgoing-wave solution of the  Regge-Wheeler equation. In the limit $r \to \infty$, this solution takes the form
 \begin{equation}
     \Psi^{\rm axial}\sim e^{i \omega x}\sum_j \alpha_j r^{-j}\, ,
 \end{equation}
with 
 \begin{align}
     \alpha_1 &= -i\frac{\ell(1+\ell)\alpha_0}{2\omega},\\
     \alpha_2 &= \left[\frac{3iM}{2\omega} + \frac{\ell(1-\ell^2)(2+\ell)}{8\omega^2}\right]\alpha_0.
 \end{align}

\item We iterate the above procedure until the interior and the exterior solution match at the star surface. The quasinormal-mode frequencies $\omega = (\omega_R, \omega_I)$ are determined by requiring the vanishing of the following expression:
 \begin{equation}
     \Delta = \psi_{\rm in}(R_s) \psi'_{\rm out}(R_s) - \psi_{\rm out}(R_s) \psi'_{\rm in}(R_s) = 0.
 \end{equation}
We implement a search algorithm in the $(\omega_R, \omega_I)$ plane to find the roots of the above expression.
 \end{itemize}

\section{Axial modes of anisotropic neutron stars}\label{sec:axialmodes}

For this work, we have focused on $\ell=2$ axial modes. As usual, we write the quasi-normal mode frequencies as $\omega = 2\pi \nu + i/\tau$, with $\nu$, $\tau$ denoting the oscillation frequency and damping time, respectively. 

Figures \ref{fig:fvsM} and \ref{fig:tauvsM} show the oscillation frequency and damping time for the base $w$-mode\footnote{The base mode refers to the lowest frequency one.} as a function of the star mass for the three EOS used and the two models of anisotropy: the Bowers-Liang model and the Horvart model.  Regardless of the stellar composition or the anisotropy prescription, the oscillation frequency generally decreases, while the damping time increases, as the mass of the configuration grows. The oscillation frequencies typically lie in the range $5$-$9$~kHz, while the damping times range from approximately $20$ to $250$~$\mu$s. Furthermore, a softer EOS typically yields slightly higher oscillation frequencies.  The influence of the anisotropy parameter $\lambda$ is equally systematic. Increasing $\lambda$ toward higher positive values leads to lower oscillation frequencies and prolonged damping times. Specifically, the Bowers-Liang model supports more massive configurations as the anisotropy parameter increases, which in turn leads to lower frequency limits and higher damping times. This behavior is consistent with the dynamical role of anisotropy within the stellar interior. Positive values of the anisotropy parameter, $\lambda$, introduce an outward force that opposes the effect of gravity. Consequently, $\lambda>0$ (i.e., tangential pressure greater than radial pressure)  allows for the equilibrium of more massive configurations compared to the isotropic case. Note that as $\lambda \to 0$, the model reduces to the isotropic case, wherein the coupling between metric and matter perturbations vanishes. In this limit, the frequency is $\sim 8$ kHz and the damping time is $\sim 50\ \mu$s, consistent with the findings of \citep{Blazquez-Salcedo:2012hdg}.

\begin{figure}
    \centering
    \includegraphics[width=0.99\linewidth]{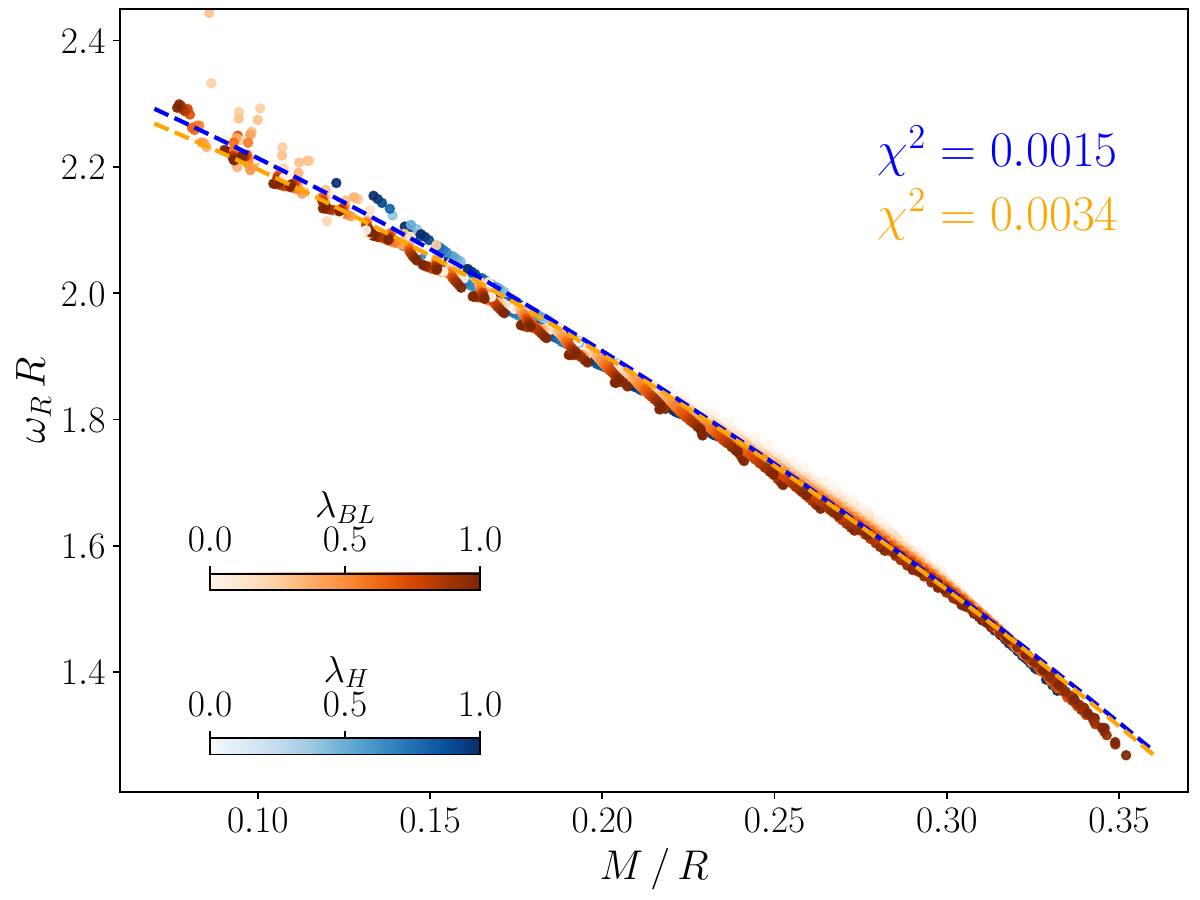}
    \includegraphics[width=0.99\linewidth]{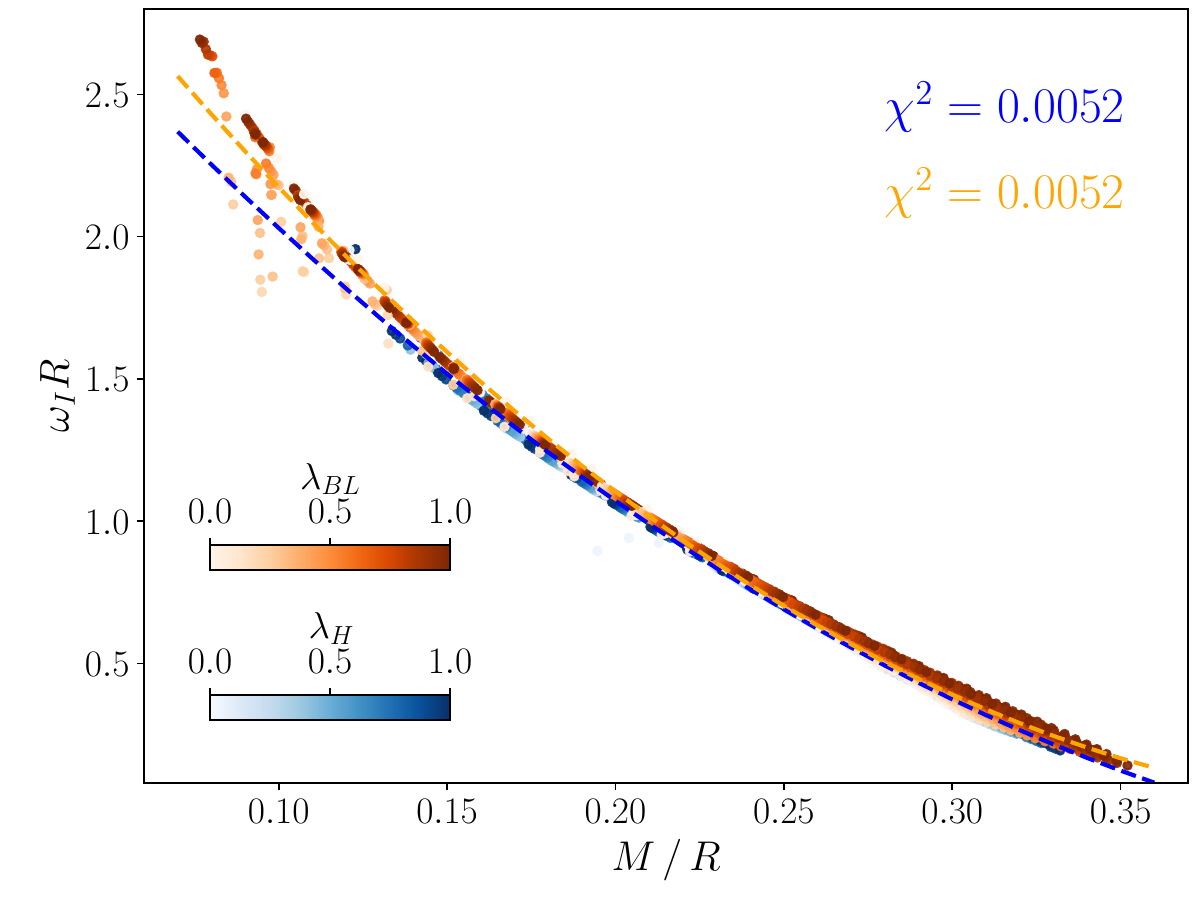}
    \caption{Oscillation frequency (upper panel) and damping time (lower panel) scaled with the stellar radius as a function of the star compactness for all three EOS used and two anisotropy models, the  Bowers-Liang (oranges color bar)and Horvart model (blues color bar). The color scale corresponds to the value of the anisotropic parameter. The dashed black line corresponds to the fit given in equation~(\ref{eq:fit_omegaR}) for the upper panel and equation~(\ref{eq:fit_omegaI}) for the lower one.}
    \label{fig:OI_MR}
\end{figure}

Figure~\ref{fig:OI_MR} shows the frequency of the base mode and the damping time scaled with the radius as a function of the star compactness, respectively. Notably, these quantities follow a nearly universal behavior with compactness, remaining largely independent of the EOS, but slightly dependent on  the anisotropy model at small compatness.

Motivated by previous studies, and to exploit future gravitational-wave observations, we derive empirical relations linking the frequency and damping time for the base $w$-mode for $\ell=2$ to the stellar compactness. To this end, we propose a quadratic fit for both quantities: 
\begin{equation}\label{eq:fit_omegaR}
     \omega_R R \approx  a_1\left(  \frac{M}{R}\right)^2+ b_1\frac{M}{R}+c_1\, ,
\end{equation}
 with $a_1^{BL}=-3.471 \pm 0.159$, $b_1^{BL} =-2.013 \pm 0.079 $ and $c_1^{BL}=2.449\pm 0.009$ for the Bowers-Liang anisotropy model, and $a_1^{H}=-3.958 \pm 0.111$, $b_1^{H} =-1.745 \pm 0.053 $ and $c_1^{H}=2.410\pm 0.005$ for the Horvart model; and
\begin{equation}\label{eq:fit_omegaI}
    \omega_I R \approx a_2\left(  \frac{M}{R}\right)^2+ b_2\frac{M}{R}+c_2\, ,
\end{equation}
 with $a^{BL}_2=13.066\pm 0.400$, $b^{BL}_2 =-13.507\pm 0.192 $ and $c^{BL}_2=3.250\pm 0.022$ for the Bowers-Liang model and $a^{H}_2=17.787\pm 0.280$, $b^{H}_2 =-16.031\pm 0.127 $ and $c^{BL}_2=3.598\pm 0.013$ for the Hovart model.

\begin{figure}
    \centering
    \includegraphics[width=0.99\linewidth]{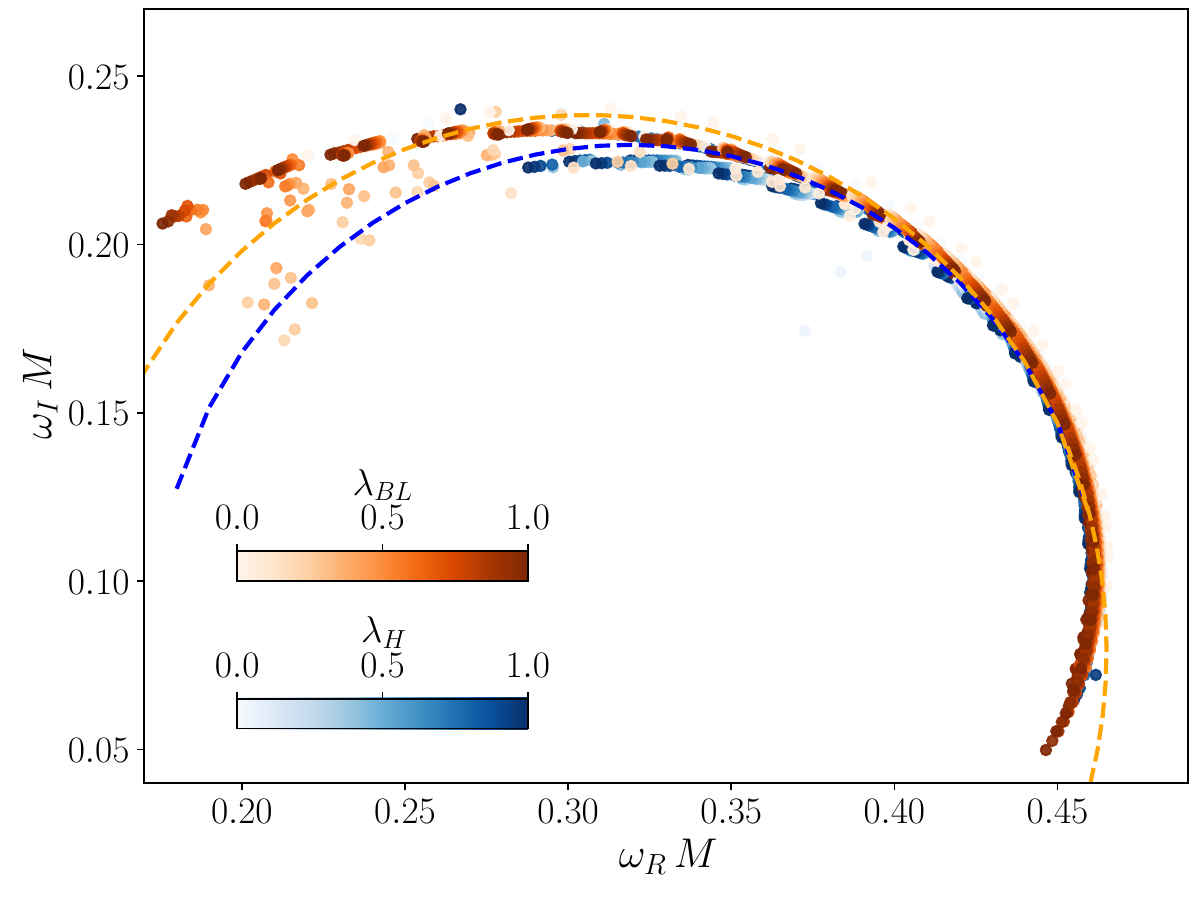}
    \caption{Damping time as a function of the oscillation frequency, both scaled by the star mass, for all three EOS used and two anisotropy models, the  Bowers-Liang  (oranges color bar)  and Horvart model (blues color bar). The color scale corresponds to the value of the anisotropic parameter. The dashed black line corresponds to the fit given in equation~(\ref{eq:fit_OROI}).}
    \label{fig:OROI}
\end{figure}

Figure~\ref{fig:OROI} shows the damping time as a function of the frequency, both scaled by the star mass. The modes lie along two distinct curves, each corresponding to one of the anisotropic models and appearing to be in independent of the EOS. These curves seem to converge as the stellar compactness increases. These  relations can be  described by the following fit:
\begin{equation}\label{eq:fit_OROI}
(\omega_R + a_3)^2 + (\omega_I+ b_3)^2 - c_3=0,
\end{equation}
with $a_3^{BL}=-0.318\pm 0.005$, $b_3^{BL} =0.0841\pm 0.0006 $ and $c_3^{BL}=0.0211\pm  0.0002$ for the Bowers-Liang anisotropy model and $a_3^{H}=-0.305\pm 0.003$, $b_3^{H} =0.0793\pm 0.0005 $ and $c_3^{H}=0.0253\pm  0.0001$ for the Hovart anisotropy model.

\section{\label{sec:discussion} Discussions and Conclusions}
In this work, we have studied the axial quasi-normal modes of anisotropic neutron stars. The stability of the axial configurations is confirmed by the positive damping times obtained through numerical integration, which ensure that the perturbations decay over time rather than grow exponentially. A foundational assumption of this analysis is that the anisotropy is an intrinsic property "frozen" into each fluid element and Lie-dragged along with its displacement. Because the spacelike vector $k^\alpha$ is geometrically defined as the unit radial vector of the background spherical symmetry, its Lagrangian perturbation, $\Delta k^\alpha$, is fully determined by the metric perturbation and the fluid displacement vector. Consequently, $\Delta k^\alpha$ is strictly constrained to lie within the $(u,k)$ plane, representing a geometric necessity to preserve the orthogonality and normalization of the perturbed system.  

This specific geometric behavior leads directly to one of the most critical physical findings of the study: for realistic anisotropic equations of state, matter and metric perturbations are intrinsically coupled for axial modes. In standard isotropic non-rotating models, axial perturbations do not involve variations in pressure or energy density, and they decouple entirely to form pure spacetime $w$-modes. However, because the anisotropic pressure variations force the fluid displacement to interact with the metric fluctuations, the quasinormal modes isolated in this work are not strictly pure $w$-modes. Instead, they are coupled modes that incorporate both gravitational and fluid degrees of freedom.  

Despite this fundamental coupling, the macroscopic behavior of these modes still strongly resembles traditional spacetime oscillations. They exhibit exceptionally short damping times in the microsecond range and sustain high oscillation frequencies spanning from approximately 5 kHz up to 9 kHz, depending on the stellar mass and the degree of anisotropy. The analysis reveals that, regardless of the underlying particle composition or the macroscopic anisotropy model utilized (Bowers-Liang or Horvart), both the scaled oscillation frequency and the scaled damping time follow a robust, universal quadratic dependence on the stellar compactness. These universal relations provide a highly reliable theoretical framework for future high-frequency gravitational-wave asteroseismology, allowing observers to probe the dense, anisotropic interiors of neutron stars directly from their coupled axial ringing.  

Finally, possible astrophysical scenarios that could excite these modes include core collapse, post-merger remnant oscillations, or pulsar glitch events. Their gravitational-wave signature would carry direct imprints of the internal pressure anisotropy absent in isotropic models, making third-generation detectors such as the Einstein Telescope or Cosmic Explorer the natural instruments to search for them.

Finally, higher frequencies are reported in \cite{Mondal:2026hki}, which assumes that $\delta k_\mu =0$ (i.e., the perturbations decouple). Our uncoupled model (see Appendix \ref{app:uncoupled}) suggests that this difference is not driven by the absence of matter-metric coupling. Instead, we conjecture that the referenced work isolates an excited mode, whereas our results represent the lowest-frequency mode.

\appendix 

\section{Equations }\label{app:I}

\subsection{Static Background Configuration}

The structure equations describing the unperturbed static background configuration are given by
\begin{eqnarray}
    \frac{rA'}{A} &=& -1+\frac{1}{B}+ \frac{8\pi r^2 \bar{P} }{B}\, ,\label{eq:da} \\
      r B' &=& 1-B-8\pi r^2\bar{\epsilon}\, , \label{eq:db}
\end{eqnarray}
while the TOV equation is
\begin{equation}
    \bar{P}'=-(\bar{P}+\bar{\epsilon})\frac{A'}{2A}-\frac{2(\bar{P}-\bar{P}_\perp)}{r} \label{eq:dP}\, .
\end{equation}
To guarantee regularity at the center, we expand all quantities around the center:
\begin{align}
    A &\approx A_c + A_2r^2+ A_4r^4,\\
B & \approx1 - \frac{8\pi\bar{\epsilon}_c}{3}r^2-\frac{8\pi\bar{\epsilon}_2}{5}r^4,\\
\bar{P} & \approx  P_c + P_2r^2+P_4r^4,
\end{align}

with
\begin{align}
    A_2 &= 4\pi\left(\bar{P}_c+\frac{\bar{\epsilon}_c}{3}\right),\\
    A_4 &= 2\pi A_c  \left[ P_2 + \frac{\epsilon_2}{5}  + 
   4 \pi\left(P_c + \bar{\epsilon}_c\right) \left( P_c + \frac{\bar{\epsilon}_c}{3}\right) \right],\\
   P_2 &=  -\frac{A_2}{2A_c}(\bar{P_c}+\bar{\epsilon}_c)-\sigma_2,\\
   P_4 &= \frac{(\bar{P}_c+\bar{\epsilon}_c)}{2}\left(\frac{A_2^2}{2A_c^2} -\frac{A_4}{A_c}\right) -\frac{\sigma_4}{2}-\frac{A_2}{4A_c}(P_2+\bar{\epsilon}_2),\\
   \epsilon_2 &= P_2 \left(\frac{d\bar{\epsilon}}{d\bar{P}}\right)_c,
\end{align}

where the subscript $c$ denotes central values. 

Finally, the power-series expansion around the center of the anisotropy is:
\begin{equation}
    \bar{\sigma}\approx \sigma_2 r^2 + \sigma_4 r^4 + \sigma_6 r^6.
\end{equation}

For the Bowers–Liang anisotropy model:
\begin{align}
    \sigma_2^{BL}&=-\lambda_{BL}(P_c+\epsilon_c)(3P_c+\epsilon_c) ,\\
    \sigma_4^{BL}&=-2\lambda_{BL}\left[\epsilon_2\left(2P_c+\epsilon_c\right) +P_2(3P_c+2\epsilon_c)+\right.\nonumber \\
    &\left.\frac{4\pi} {3}\epsilon_c(3P_c^2+4P_c\epsilon_c+\epsilon_c^2)\right],
\end{align}

and for the Horvart model:

\begin{align}
    \sigma_{2}^H&=0,\\
    \sigma_4^{H}&=-\lambda_H \left(\frac{ 8\pi\epsilon_c}{3}\right)^2P_c,\\
    \sigma_6^{H}&=-\lambda_H \left(\frac{ 8\pi\epsilon_c}{3}\right)^2P_c\left(\frac{6}{5}\frac{\epsilon_2}{\epsilon_c}+\frac{P_2}{P_c}\right).
\end{align}


\subsection{Axial Perturbations Equations}

First, in order to improve the numerical behavior, and assuming that the variables have an harmonic time dependence we perform the following change of variables,  
\begin{align}
    h_0 \rightarrow & i \omega r^{\ell +1} h_0(r) e^{i \omega t},\\
    h_1 \rightarrow & r^{\ell +2} h_1(r) e^{i \omega t},\\
    X \rightarrow & r^{\ell - 1} X(r) e^{i \omega t}.
\end{align}

The system of equations that describe the axial perturbation of anisotropic configurations is, 
\begin{eqnarray}
    rh_1'&=& -h_0\frac{\omega^2 }{AB} -rh_1\left( \frac{2+\ell}{r}+\frac{A'}{2A}+\frac{B'}{2B}\right),\label{eq:axiala}\\
    r h_0' &=& h_0(1-\ell)+h_1\left(r^2-(\ell^2+\ell-2)\frac{A}{\omega^2}\right)  \nonumber \\ & & +\frac{ 16\pi A}{\omega^2}\chi,\label{eq:axialb} \\
     rX'&=& -r^2h_1+X(1-\ell) +\frac{\chi}{\bar{\sigma}},\label{eq:axialc}\\
    \chi'&=& \left(X +h_0\right)\frac{r(\bar{P}+\bar{\epsilon}-\bar{\sigma})\omega^2}{AB}\label{eq:axiald} \\
    & &-  \chi\left(\frac{2+\ell}{r} +\frac{A'}{2A}+\frac{B'}{2B}\right),\nonumber
\end{eqnarray}
where we have defined $\chi = \bar{\sigma} [ (\ell-1)X + rX'+ r^2h_1 ]$.

Since the equations are singular at the origin, the integration is initiated at a small radius 
$\Delta r$. The boundary conditions are then obtained by expanding the variables  in powers of $r$ around the origin,
\begin{align}
    h_0 &\approx h_{0,c}+h_{0,2}r^2,\\
    h_1 &\approx h_{1,c}+h_{1,2}r^2,\\
    X & \approx X_c+X_2 r^2, \\
    \chi &\approx \chi_{2}r^2 + \chi_4 r^4\, .
\end{align}
After replacing the series expansion in Eqs. (\ref{eq:axiala})-(\ref{eq:axiald}), the coefficients are

\begin{align}
    h_{1,c} &= -\frac{\omega^2h_{0,c}}{A_c(2+l)}\, , \\
     X_c &= \frac{(\bar{P}_c+\bar{\epsilon}_c)\omega^2h_{0,c}}{A_c(\ell-1)(\ell+4)\sigma_2-(\bar{P}_c+\bar{\epsilon}_c)\omega^2}, \\
    \chi_2 &= (\ell-1)\sigma_2 X_c,\\
    h_{0,2} &=\frac{3+\ell}{2+\ell}\left(B_2+\frac{A_2}{A_c}\right)h_{0,c}-\frac{A_c(4+\ell)}{\omega^2}h_{1,2},\\
    h_{1,2}&=-\left[\frac{B_2A_c(1+\ell)(3+\ell)+ A_2(5+3\ell) +\omega^2}{2(3+2\ell)A_c}\right]h_{1,c} \nonumber \\
    & -\frac{8\pi \chi_2}{3+2\ell},\\
    \chi_4 &= \frac{\sigma_2\tilde{\chi}_4}{A_c(1+\ell)(6+\ell)\sigma_2-(\bar{P}_c+\bar{\epsilon}_c)\omega^2},\\
    X_2 &= \frac{1}{1+\ell}\left(-h_{1,c} + \frac{\chi_4}{\sigma_2}  - \chi_2\frac{\sigma_4}{\sigma_2}\right),
\end{align}
with,

\begin{multline}
    \tilde{\chi}_4= (\bar{P}_c+\bar{\epsilon}_c)\left[-h_{1,c}+h_{0,2}(\ell+1)+X_c(\ell-1)\frac{\sigma_4}{\sigma_2}\right]\omega^2\\+(\ell+1)\left( \frac{A_2}{A_c} + B_2 \right)\left[X_c(\ell-1)A_c\sigma_2 -(h_{0,c}+X_c)(\bar{P}_c+\bar{\epsilon}_c)\omega^2\right]\\ +(\ell+1) \left(h_{0,c}+X_c\right)\left(P_2+\epsilon_2-\sigma_2\right)\omega^2
\end{multline}

\section{Metric Perturbation decoupled} \label{app:uncoupled}

\begin{figure}[t!]
    \centering
    \includegraphics[width=0.85\linewidth]{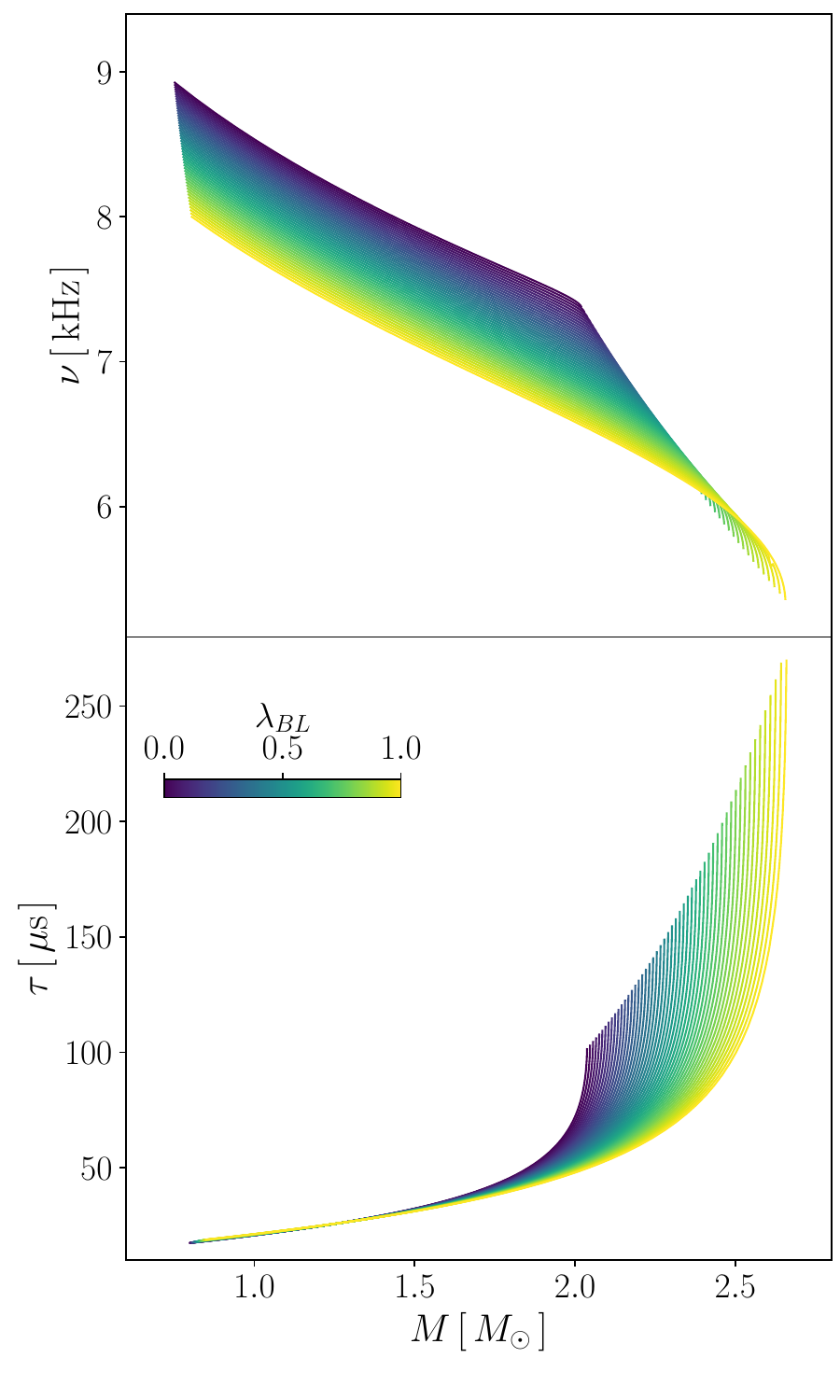}
    \caption{Oscillation frequency (top panel) and damping time (bottom panel) of the base $w$-mode as a function of the stellar mass for the SLy4 EOS and the Bowers-Liang anisotropy model.}
    \label{fig:Omegadk0}
\end{figure}
Setting $(\delta k^\alpha)^{\mathrm{axial}}=0$ causes the metric perturbations to decouple from the fluid displacement. The axial modes are then described by Eqs.~\eqref{eq:axiala} and \eqref{eq:axialb} with $\chi=0$. The procedure used to integrate the equations and determine the base mode is identical to that described in Section~\ref{sec:numerical}.

Figure~\ref{fig:Omegadk0} shows the oscillation frequency and damping time of the base $w$-mode as functions of the stellar mass for the SLy4 EOS and the Bowers--Liang anisotropy model. Table~\ref{tab:Omegadk0} compares the oscillation frequencies and damping times obtained for $(\delta k^\alpha)^{\mathrm{axial}}=0$ and $(\delta k^\alpha)^{\mathrm{axial}}\neq 0$, where the latter is given by Eq.~\eqref{eqn:Deltak}. The comparison is performed for three values of the anisotropy parameter and several central mass densities. The differences remain below $10\%$, even for the largest anisotropies considered.
\begin{table*}[]
    \centering
  {  \begin{tabular}{c|ccccccccc}   
     & $ \log_{10}(\rho_c)$ & $M/R$ & $\nu (\delta k= 0)$ &$\nu (\delta k\neq0)$&$\Delta \nu$ & $\tau (\delta k= 0)$& $\tau (\delta k\neq  0)$  & $\Delta \tau$ \\
     &  & & kHz & kHz & kHz & $\mu$s & $\mu$s & $\mu$s \\ \hline
       & $14.851 $ &  $0.138 $ & $8.272$ & $8.511$ & $ -0.239$ &  $23.063$ & $24.702$  & $-1.639$ \\
    & $14.990$ & $0.203$  & $7.656$ & $7.654$  &$0.0021$ & $36.610$  & $37.722$ &  $-1.112$ \\
   $\lambda_{BL}=0.2$   & $15.096$ & $0.244$  &$7.334$& $7.294$ & $0.039$ &$53.559$ &  $54.614$   & $-1.055$ \\
    & $15.180$ & $0.267$  & $7.152$&  $7.118$ & $0.033$  & $75.356$ & $75.721$ &$-0.365$\\
    & $15.251$ &  $ 0.281$ & $7.037$& $7.021$ & $0.016$&  $104.419$ & $103.592$  & $0.826$ \\ \hline 

   &  $14.813$&  $0.138$& $7.997$ &$8.103$  &$-0.106$ & $23.756$  &  $24.271$ & $-0.515$ \\
     & $14.952$ & $0.208$ & $7.318$& $7.316$ & $0.001$ & $38.999$  & $40.163$  & $-1.164$ \\
 $\lambda_{BL}=0.5$   & $15.057$ &  $0.252$&$6.924$ & $6.890$ & $0.033$&$59.667$& $60.603$ & $-0.935$    \\
& $15.142$ & $0.278$ &$6.669$ & $6.649$ &$0.0199$ &  $88.751$ &$88.499$   &$0.251$  \\
& $15.213$ & $0.293$ & $6.480$&  $6.489$& $-0.009$& $131.900$  &  $129.262$ & $2.637$ \\ \hline

  & $14.735$  & $0.137$ & $7.474$ &$7.586$ & $-0.112$ &$25.207$  & $25.752$  & $-0.545$ \\
  & $14.875$  & $0.214$ & $6.711$& $6.714$ & $-0.003$ & $43.606$& $44.93$ &  $-1.327$   \\
 $\lambda_{BL}=1.0$  & $14.980$ & $0.267$ & $6.184$& $6.162$ & $0.021$&  $73.343$ & $74.265$  &$-0.922$  \\
 & $15.065$ & $0.298$ & $5.794$&  $5.797$&  $-0.003$ & $123.396$  &  $ 122.170$  &$1.226$  \\
 & $15.136$ & $0.315$ &$5.471$ &$5.513$  &$-0.041$ & $216.416$  &$215.103$   & $1.313$ \\ \hline
    \end{tabular}}
    \caption{Comparison of the oscillation frequency  and damping time  of the base $w$-mode for the SLy4 EOS and the Bowers-Liang anisotropy model.}
    \label{tab:Omegadk0}
\end{table*}


\begin{acknowledgments}
We thank the anonymous referee for their valuable comments and suggestions. 
 J.F.R.R is supported by the Universidad Antonio Nari\~no. F.D.L-C is supported by the Vicerrectoría de Investigación y Extensión - Universidad Industrial de Santander.
\end{acknowledgments} 

\section*{DATA AVAILABILITY}
The data that support the findings of this article are not publicly available upon publication because it is not technically feasible and/or the cost of preparing, depositing, and hosting the data would be prohibitive within the terms of this research project. The data are available from the authors upon reasonable request.


\bibliography{apssamp}

\end{document}